\documentstyle[aps,prl]{revtex}
\input epsf
\input{epsf.sty}
\begin{document}
\draft
\twocolumn[\hsize\textwidth\columnwidth\hsize\csname@twocolumnfalse\endcsname

\title {Abrupt Change of 
Josephson Plasma Frequency at the Phase Boundary of\\
the Bragg Glass in Bi$_2$Sr$_2$CaCu$_2$O$_{8+\delta}$} 
\author{M.B. Gaifullin$^1$, Yuji Matsuda$^{1,\dag}$, N. Chikumoto$^2$, J. 
Shimoyama$^3$ and K. Kishio$^3$} \address{$^1$Institute for Solid State Physics, 
University of Tokyo, Roppongi 7-22-1, Minato-ku, Tokyo 106, Japan} \address{$^2$Superconductivity 
Research Laboratory, ISTEC, Shibaura 1-16-25 , Minato-ku, Tokyo 105, Japan} 
\address{$^3$Department of Superconductivity, University of Tokyo, Bunkyo-ku, 
Tokyo 113, Japan}

\date{received 16 December 1999} 
\maketitle
\begin{abstract}

We report the first detailed and quantitative study of the Josephson coupling 
energy in the vortex liquid, Bragg glass and vortex glass phases of 
Bi$_2$Sr$_2$CaCu$_2$O$_{8+\delta}$ by the Josephson plasma resonance.  The 
measurements revealed distinct features in the $T$- and $H$-dependencies of the 
plasma frequency $\omega_{pl}$ for each of these three vortex phases.  When 
going across either the Bragg-to-vortex glass or the Bragg-to-liquid transition 
line, $\omega_{pl}$ shows a dramatic change.  We provide a quantitative 
discussion on the properties of these phase transitions, including the first 
order nature of the Bragg-to-vortex glass transition.

\end{abstract}
\pacs{74.25.Nf, 74.50+r, 74.60.Ec, 74.72.Hs}

]

\narrowtext

The vortex matter in 
high-$T_c$ superconductors exhibits a fascinatingly rich phase diagram with a 
variety of phase transitions.  There, thermal fluctuation and disorder alter 
dramatically the vortex phase diagram which has been observed in the 
conventional superconductors.  At high temperature, the strong thermal 
fluctuation melts a vortex lattice into a vortex liquid well below the upper 
critical field.  On the other hand, at low temperature or low field where the 
vortex liquid freezes into a solid phase, disorder plays an important role.  The 
disorder is known to destroy the long-range order of the Abrikosov lattice 
\cite{lo}.  Recent investigations have revealed that the vortex solid phase is 
comprised of two distinct phases; a highly disordered phase at high field and a 
rather ordered phase at low field \cite{chiku,cubt,khay}.  The former phase is 
the vortex glass or entangled solid phase which is characterized by divergent 
barriers for vortex motion \cite{ffh}.  The latter phase is the Bragg glass or 
quasilattice phase in which no dislocation exists and quasi-long-range 
translational order is preserved \cite{giam}.  In very clean single crystals, 
thermodynamical measurements have revealed that the Bragg glass undergoes a 
first order transition (FOT) to the vortex liquid \cite{zeld,dods,hu}.  The 
transition from the Bragg glass to the vortex glass, on the other hand, is 
characterized by the second magnetization peak at which the critical current 
shows a sharp increase \cite{khay}.  It was proposed that the crossover from the 
FOT to the second peak regime is governed by a critical point $T_{cp}$ in the 
phase diagram, which in Bi$_2$Sr$_2$CaCu$_2$O$_{8+\delta}$ is located near 40~K. 
While the nature of the vortex liquid has been extensively studied, the 
properties of the Bragg glass and the nature of the thermally induced FOT from 
the Bragg glass to the vortex liquid are still not quite understood.  Moreover, 
the phase transition from the Bragg glass to the vortex glass at lower 
temperatures has been a longstanding issue, though this transition is proposed 
to be disorder driven, caused by competition between the elastic and pinning 
energies \cite{khay,nels,ryu,vino,horo,kosh2}.  A major obstacle has been that 
most of the previous experiments had to been performed under a strongly 
nonequilibrium condition because most part of the Bragg and vortex glasses are 
located deep inside the irreversibility line $T_{irr}$.

The most direct way to clarify the nature of these phases and the phase 
transitions among them is to measure the interlayer phase coherence for each 
vortex phases, because the CuO$_2$ layers are connected by the Josephson effect.  
One of the most powerful probes for the interlayer phase coherence is the 
Josephson plasma resonance (JPR) which provides a direct measurement of the 
Josephson plasma frequency $\omega_{pl}$ related to the maximum Josephson 
critical current $j_{J}=\varepsilon_0\Phi_0\omega_{pl}^2/8\pi^2cd$ and the 
Josephson coupling energy $U_J=\Phi_0j_J/2 \pi c$, where $\varepsilon_0$ and $d$ 
are the dielectric constant and interlayer spacing, respectively.  
\cite{mats1,mats2,shib,bula,kosh1,hwan,gaif,kosh3}.  Especially in 
Bi$_2$Sr$_2$CaCu$_2$O$_{8+\delta}$ with large anisotropy, a very precise 
determination of $\omega_{pl}$ is possible because $\omega_{pl}$ falls within 
the microwave window.

All of the JPR measurements of Bi$_2$Sr$_2$CaCu$_2$O$_{8+\delta}$ up to now have 
been carried out in the cavity resonator by reducing $\omega_{pl}$ by $H$ 
\cite{mats1,mats2,shib}.  Unfortunately, sweeping $H$ below $T_{irr}$ drives the 
vortex system into a strongly nonequilibrium state due to the Bean critical 
current induced by the field gradient inside the crystal, as was demonstrated in 
Refs.\cite{mats2} and \cite{rodr}.  Therefore, in order to investigate the Bragg 
and vortex glass phases, it is crucial to measure the JPR as a function of the 
microwave frequency $\omega_{pl}$ while holding $H$ at a constant value.  In 
this Letter, we report the first detailed and quantitative study of the 
Josephson coupling energy in the Bragg glass, the vortex glass and the vortex 
liquid phases and the phase transitions among them by the JPR which has been 
preformed by sweeping $\omega$ continuously.  The measurements revealed distinct 
features in the $T$- and $H$-dependencies of $\omega_{pl}$ for each of the three 
different vortex phases.  When going across either the Bragg-to-vortex glass

\begin{figure}
\centerline{\epsfxsize 7.5cm \epsfbox{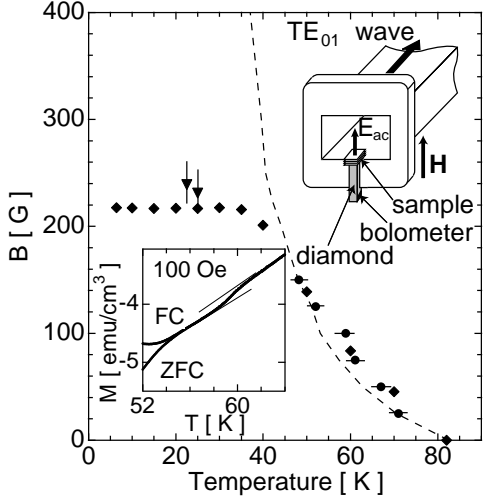}}
\vspace{-1cm}
\caption{ Vortex phase diagram determined by 
the magnetization and the JPR. The filled circles represent $H_{m}$ determined 
by the magnetization step (see inset).  The filled triangles represent $H_{sp}$.  
The filled diamonds represent the field at which $\omega_{pl}$ shows an abrupt 
change.  The dashed line is the irreversibility line.  Lower inset: 
Magnetization measured in the field cooling (FC) and zero field cooling (ZFC) 
conditions.  Upper inset: Bolometric detection of the microwave absorption.  The 
crystal is supported inside the waveguide by a thermally isolated diamond plank 
({\bf H}$\parallel c$).  The JPR is caused by the microwave electric field {\bf 
E}$_{ac}\parallel c$.}

\end{figure}
\noindent or 
the Bragg-to-liquid transition line, $\omega_{pl}$ shows a dramatic change.  We 
provide a quantitative discussion on the nature of these phase transitions in 
the light of these results.

All experiments were performed on a slightly underdoped 
Bi$_2$Sr$_2$CaCu$_2$O$_{8+\delta}$ single crystals ($T_c$=82.5~K) with 
dimensions $1.2\times0.5\times0.03$mm$^3$ grown by the traveling floating zone 
method.  The inset of Fig.1 shows a typical magnetization step measured by SQUID 
magnetometer which can be attributed to the FOT of the vortex lattice.  This FOT 
terminates at $\approx40$~K and the step is followed by the second magnetization 
peak located at $\sim$230~Oe.  Figure 1 shows the phase diagram obtained by the 
magnetization measurements.  The JPR is measured by sweeping $\omega$ 
continuously from 20~GHz to 150~GHz \cite{gaif}.  The sample was placed at the 
center of the broad wall of the waveguide in the traveling wave TE$_{01}$ mode.  
We used a bolometric technique to detect very small microwave absorption by the 
sample and employed a leveling loop technique to ckeep the microwave power 
constant when sweeping frequency.  For this crystal $\omega_{pl}$=125~GHz at 
$T$=0, corresponding to the anisotropy parameter 
$\gamma=\lambda_c/\lambda_{ab}\approx 550$, where $\lambda_{ab}$ and $\lambda_c$ 
are the in-plane and out-of-plane penetration lengths, respectively.  Here we 
used $\lambda_{ab}\approx 200$~nm and 
$\lambda_c=c/\omega_{pl}\sqrt{\varepsilon_0}\approx$110~$\mu$m.  We determined 
$\varepsilon_0=11.5\pm1$ from the dispersion of the transverse plasma mode.  All

\begin{figure}
\centerline{\epsfxsize 7.5cm \epsfbox{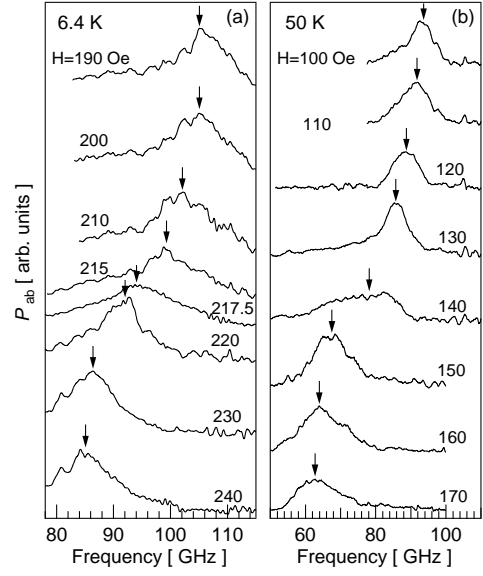}}
\caption{The JPR as a function of frequency 
when crossing (a) the second peak (6.4~K) and (b) the FOT (50~K).  We picked up 
only the longitudinal plasma mode which is sample size independent.  The arrows 
indicate the peak position. }
\end{figure}

\noindent JPR measurements were performed in {\bf H}$\parallel c$ {\em under the field 
cooling condition} (FCC) where the field is very uniform.  In this condition, 
the system is in equilibrium or at worst is trapped in a metastable state which 
we expect should be much closer to equilibrium compared to the state obtained in 
the field sweeping condition (FSC).  In fact, while the resonance frequency 
below $T_{irr}$ did not change at all with time for more than 48 hours in the 
FCC, it increases gradually with time in the FSC. We also confirmed that the 
resonance curves are exactly the same in different cooling cycles.

Figures 2(a) and (b) depict the resonant absorption as a function of 
$\omega$ when crossing the second peak field $H_{sp}$ and the FOT field 
$H_m$, respectively.  When $\omega$ coincides with $\omega_{pl}$, the 
resonant absorption of the microwave occurs.  These are the JPR measured 
in the Bragg and vortex glass phases under the FCC for the first time.  
In the magnetic field, $\omega_{pl}$ can be written as \cite{bula}
\begin{equation}
\omega_{pl}^2(B,T)=\omega_{pl}^2(0,T)\langle\cos\phi_{n,n+1}\rangle.
\end{equation}
Here $\langle\cos\phi_{n,n+1}\rangle$ represents the thermal and disorder 
average of the cosine of the gauge invariant phase difference between layer $n$ 
and $n+1$.  If the vortex forms a straight line along the $c$-axis, 
$\langle\cos\phi_{n,n+1}\rangle$ is unity.  The reduction of 
$\langle\cos\phi_{n,n+1}\rangle$ from unity is caused by the Josephson strings 
that are created by the deviation from the straight alignment of the pancake 
vortices along the $c$-axis.  Thus $\omega_{pl}$ gives a direct information on 
the vortex alignment and therefore the phase transition of the vortex matter.  
After gradual decrease with $H$ at lower $H$, $\omega_{pl}$ shows a sharp 
decrease in the field range between 215~Oe and 220~Oe at 6.5~K and between 
140~0e and 

\begin{figure}
\centerline{\epsfxsize 7.5cm \epsfbox{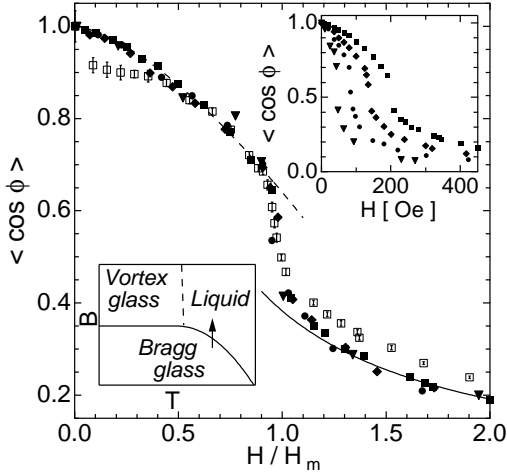}}
\caption{Inset: $H$-dependence of 
$\langle\cos\phi_{n,n+1}\rangle$ when going across the FOT at high temperatures.  
Solid squares, diamonds, circles, and triangles show the data at 40~K, 50~K, 
60~K, and 70~K respectively.  Main panel: Same data plotted as a function of 
$H/H_m$.  Open squares show $\langle\cos\phi_{n,n+1}\rangle$ as a function of 
$H/H_{sp}$ at 30~K when crossing the second peak.  The dashed line is the result 
of Eq.2.  The solid line is the fit to Eq.3.  }
\end{figure}
\noindent 160~Oe at 50~K. At 217.5~Oe in Fig.2(a) and at 150~Oe in Fig.2(b), 
the resonance lines become broader, indicating a very rapid change of 
$\omega_{pl}$ with $H$.  At higher $H$, $\omega_{pl}$ again decreases gradually.  
In Fig.1, we plot the fields at which $\omega_{pl}$ shows an abrupt change.  
These fields coincide well with the second peak and FOT fields determined by 
magnetization measurements.

We first discuss the resonance when going across the FOT. The inset of Fig.3 
depicts the $H$-dependence of $\langle\cos\phi_{n,n+1}\rangle$ obtained from 
$\omega_{pl}^2(B,T)/\omega_{pl}^2(0,T)$.  Although similar results have been 
reported \cite{mats2,shib}, quantitative analysis was very difficult because the 
JPR measurements in the Bragg glass had been done under the strongly 
nonequilibrium condition, as we have already mentioned.  Figure 3 depicts 
$\langle\cos\phi_{n,n+1}\rangle$ as a function of $H$ normalized by $H_m$.  
Interestingly, $\langle\cos\phi_{n,n+1}\rangle$ exhibits very similar 
$H/H_m$-dependence at all temperatures.  Obviously, the $H$-dependence of 
$\langle\cos\phi_{n,n+1}\rangle$ above FOT is very different from that below 
FOT; the curvature changes from negative to positive.  We found that 
$\langle\cos\phi_{n,n+1}\rangle$ in the Bragg glass phase can be fitted as,
\begin{equation}
\langle\cos\phi_{n,n+1}\rangle= 
1-A_1\frac{H}{H_m}-A_2\left(\frac{H}{H_m}\right)^2,
\end{equation}
with $A_1$=0.16 and $A_2$=0.19 above 40~K as shown in the dashed line in Fig.3.  
On the other hand, according to high temperature expansion theory \cite{kosh1}, 
$\langle\cos\phi_{n,n+1}\rangle$ in the liquid phase above FOT can be written 
as,
\begin{equation}
\langle\cos\phi_{n,n+1}\rangle=\frac{U_J\Phi_0}{2k_BTH},
\end{equation}

\begin{figure}
 \centerline{\epsfxsize 7.5cm \epsfbox{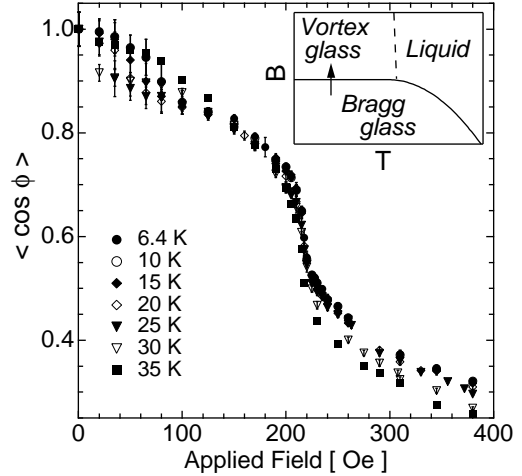}}
 \caption{$H$-dependence of 
 $\langle\cos\phi_{n,n+1}\rangle$ when going across the Bragg-to-vortex glass 
 transition at low temperatures.  }
\end{figure}

\noindent when the Josephson energy is negligible compared with the energy of thermal 
fluctuations, {\it i.e.}$U_J \ll k_BTH/\Phi_0$.  It has been shown 
experimentally that $\langle\cos\phi_{n,n+1}\rangle$ is inversely proportional 
to $H$ in the liquid phase \cite{mats1,mats2,shib}.  The present results provide 
a further rigorous test to Eq.3, because we now have no ambiguous fitting 
parameter and also have the data of the very detailed $H$-dependence of 
$\langle\cos\phi_{n,n+1}\rangle$ obtained by sweeping $\omega$.  The solid line 
in Fig.3 shows the result of the calculation.  In the calculation, we used 
$\varepsilon_0$=12.0.  The fit to the data is excellent
in the whole $H$-range at $H>1.2H_m$, indicating that the vortex 
liquid is decoupled on the scale of the interlayer distance.  Small deviation 
from Eq.3 is observed at $H\leq1.2H_m$.  This suggests that the vortex-vortex 
correlation effect in the $ab$-plane which gives rise to the deviation from 
$1/H$-dependence of $\langle\cos\phi_{n,n+1}\rangle$ in the liquid phase is 
important just above the FOT \cite{kosh3}.

At $H_m$, $\langle\cos\phi_{n,n+1}\rangle$ is reduced to $\approx$0.7 at all 
temperatures, showing {\em an occurrence of large vortex wandering in the Bragg 
glass.} Near $T_c$, we note that $\omega_{pl}$ at $H$=0 is already suppressed by 
the phase fluctuations.  If this effect is taken into account, it is expected 
that $\langle\cos\phi_{n,n+1}\rangle$ at $H_m$ slowly increases with $T$, 
indicating that the melting becomes more linelike at higher $T$.  The values of 
$\langle\cos\phi_{n,n+1}\rangle$ at $H_m$ are close to the recent results of 
computer simulations for systems with small anisotropies\cite{hu}.

The internal energy $U$ experiences a jump $\Delta U$ at the FOT. This latent 
heat $\Delta U$ can be represented as a sum of the jumps in the in-plane energy, 
in the electromagnetic coupling energy, and in the Josephson energy $\Delta U_J$ 
\cite{hu}.  To understand the nature of the FOT in detail, it is important to 
establish the relative jump in Josephson energy $\Delta U_J/\Delta U$.  At 60~K, 
$\Delta U_J/T$ can be estimated to be $\approx 0.21k_B$ from 
$\langle\cos\phi_{n,n+1}\rangle$ which drops approximately from 0.70 to 0.45.  
On the other hand, $\Delta U/T$ at 60~K obtained from the magnetization step 
$\Delta M$ using the Clausius-Clapeyron relation,
\begin{equation}
\Delta U/T_m=\Delta S=-d\Phi_0\frac{\Delta M}{B_m}\frac{dB_m}{dT},
\end{equation}
is $\approx1.34k_B$.  Here $\Delta S$ is the entropy jump at the FOT point 
($T_m$, $B_m$).  Thus we find that $\Delta U_J$ constitutes approximately 16\% 
of the latent heat, showing that $\Delta U_J$ occupies a substantial part of the 
latent heat at the FOT even in Bi$_2$Sr$_2$CaCu$_2$O$_{8+\delta}$ with very 
large anisotropy.

We now move on to the subject of the Josephson coupling at low temperatures when 
going across the transition from the Bragg glass to the vortex glass.  Figure 4 
shows the $H$-dependence of $\langle\cos\phi_{n,n+1}\rangle$ below 35~K. Below 
100~Oe, $\langle\cos\phi_{n,n+1}\rangle$ shows a hump structure which may be 
related with the lower critical field.  Above 100~Oe, the $H$-dependence of 
$\langle\cos\phi_{n,n+1}\rangle$ is very similar to that at high $T$ when 
crossing the FOT. At all temperatures, $\langle\cos\phi_{n,n+1}\rangle$ shows an 
abrupt change at the second peak field $H_{sp}\approx$220~Oe.  In similarity to 
the high temperature behavior, the $H$-dependence of 
$\langle\cos\phi_{n,n+1}\rangle$ below and above $H_{sp}$ are very different, 
showing clearly that $H_{sp}$ separates two distinct vortex phases.  In a very 
narrow field interval less than 5~Oe at $H_{sp}$, 
$\langle\cos\phi_{n,n+1}\rangle$ drops from approximately 0.7 to 0.5 (see also 
Fig.1(a)), corresponding to a nearly 20\%-reduction of $U_J$.  This strong 
reduction of $\langle\cos\phi_{n,n+1}\rangle$ provides a direct evidence of the 
decoupling nature of the Bragg-to-vortex glass transition \cite{horo}.  At 
$H_{sp}$, $\langle\cos\phi_{n,n+1}\rangle$ is reduced to $\approx$0.7 from the 
zero field value similar to that below FOT. Although we do not show here, 
$\langle\cos\phi_{n,n+1}\rangle$ in the vortex glass phase deviates from the 
$1/H$ dependence in the whole $B$-regime above $H_{sp}$, which is to be 
contrasted to the behavior in the vortex liquid phase.

We finally discuss the phase transition from the Bragg glass to the vortex glass 
inferred from the JPR. The first question is the order of the transition.  The 
abrupt change of $\langle\cos\phi_{n,n+1}\rangle$ shown in Figs.3 and 4 provides 
a direct evidence of the abrupt changes of the $c$-axis correlation length of 
the pancakes and of $U_{J}$ which composes a substantial part in the free 
energy.  In Fig.3 we plot the change of $\langle\cos\phi_{n,n+1}\rangle$ at 
$H_{sp}$ ($T$=30~K), for the comparison with the change of the same quantity at 
the FOT. {\em The change of $\langle\cos\phi_{n,n+1}\rangle$ at $H_{sp}$ is 
comparable or even sharper than that at the FOT}.  This fact strongly indicates 
the first order nature of the phase transition from the Bragg glass to the 
vortex glass.  The second issue is the critical point $T_{cp}$ of the FOT which 
has been proposed to terminate at $\approx$40~K \cite{khay,zeld}.  This proposal 
was made from the observation that $\Delta S$ becomes extremely small which can 
be seen from the $T$-independence of FOT line below $T_{cp}$.  However, the 
vanishing of $\Delta S$ does not immediately imply the termination of the FOT, 
which suggests that the issue of the termination is nontrivial.  As seen in 
Figs.3 and 4, there is no discernible difference in the $H$-dependence of 
$\langle\cos\phi_{n,n+1}\rangle$ as we go through the Bragg-to-liquid transition 
regime, into the Bragg-to-vortex glass transition regime, except for a gradual 
decrease of the change of $\langle\cos\phi_{n,n+1}\rangle$ at the transition.  
These results imply that the FOT does not terminate at $\approx$40~K, but that 
there is no critical point or the FOT persists at least below 6.4~K. We note 
that a similar conclusion has been reached very recently using the 
magneto-optical imaging technique \cite{beek}.

In summary, we have performed the JPR experiments in the Bragg glass, vortex 
glass, and vortex liquid phases in the FCC. We found an abrupt change in the 
Josephson coupling energy when going across either the FOT line or the second 
magnetization peak line.  We showed that this change occupies a substantial part 
of the latent heat at the FOT. The results suggest that the Bragg-to-vortex 
glass transition is first order in nature and that the critical point of the FOT 
does not terminate at $\approx$40~K.

We thank B. Horovitz, X. Hu, Y. Kato, P.H. Kes, T. Onogi, A. Sudb\o, and A. 
Tanaka for discussions.  We are indebted to L.N. Bulaevskii
for several valuable comments. We also thank A.E.Koshelev for the critical 
reading of the manuscript.

\end{document}